\documentclass[10pt,conference]{IEEEtran}
\IEEEoverridecommandlockouts
\usepackage{cite}
\usepackage{amsmath,amssymb,amsfonts}
\usepackage{algorithmic}
\usepackage{graphicx}
\usepackage{textcomp}
\usepackage[x11names]{xcolor}
\def\BibTeX{{\rm B\kern-.05em{\sc i\kern-.025em b}\kern-.08em
    T\kern-.1667em\lower.7ex\hbox{E}\kern-.125emX}}

\usepackage{hyperref}
\usepackage{multirow}
\usepackage{booktabs}
\usepackage{caption}
\usepackage{subcaption}

\usepackage{quantikz}
\usepackage{xspace}

\usepackage{soul}

\usepackage{tikz}
\usetikzlibrary{calc,fit,positioning,shapes.multipart}

\newcommand{\bea}{\begin{eqnarray}} 
\newcommand{\eea}{\end{eqnarray}}
\newcommand{\mathbbm}[1]{\text{\usefont{U}{bbm}{m}{n}#1}} 

\newcommand{\ie}{i.e.\xspace}
\newcommand{\wrt}{w.r.t.\xspace}
\newcommand{\figref}[1]{Fig. \ref{#1}\xspace}
\newcommand{\secref}[1]{Section \ref{#1}\xspace}
\newcommand{\tabref}[1]{Table \ref{#1}\xspace}

\newcommand\orcidicon[1]{\href{https://orcid.org/#1}{\mbox{\scalerel*{
\begin{tikzpicture}[yscale=-1,transform shape]
\pic{orcidlogo};
\end{tikzpicture}
}{|}}}}

\begin{document}

\title{Reinforcement Learning for Variational Quantum Circuits Design\\
}

\author{
\IEEEauthorblockN{Simone Foderà}
\IEEEauthorblockA{\textit{Politecnico di Milano}\\
Milano, Italy\\
\href{mailto:simone.fodera@mail.polimi.it}{simone.fodera@mail.polimi.it}}
\and
\IEEEauthorblockN{Gloria Turati}
\IEEEauthorblockA{\textit{Politecnico di Milano}\\
Milano, Italy\\
\href{mailto:gloria.turati@polimi.it}{gloria.turati@polimi.it}}
\and
\IEEEauthorblockN{Riccardo Nembrini}
\IEEEauthorblockA{\textit{Politecnico di Milano}\\
Milano, Italy\\
\href{mailto:riccardo.nembrini@polimi.it}{riccardo.nembrini@polimi.it}}
\and
\IEEEauthorblockN{Maurizio Ferrari Dacrema}
\IEEEauthorblockA{\textit{Politecnico di Milano}\\
Milano, Italy\\
\href{mailto:maurizio.ferrari@polimi.it}{maurizio.ferrari@polimi.it}}
\and
\IEEEauthorblockN{Paolo Cremonesi}
\IEEEauthorblockA{\textit{Politecnico di Milano}\\
Milano, Italy\\
\href{mailto:paolo.cremonesi@polimi.it}{paolo.cremonesi@polimi.it}}
}

\maketitle

\thispagestyle{plain}
\pagestyle{plain}

\begin{abstract}
Variational Quantum Algorithms have emerged as promising tools for solving optimization problems on quantum computers.
These algorithms leverage a parametric quantum circuit called ansatz, where its parameters are adjusted by a classical optimizer with the goal of optimizing a certain cost function.
However, a significant challenge lies in designing effective circuits for addressing specific problems.
In this study, we leverage the powerful and flexible Reinforcement Learning paradigm to train an agent capable of autonomously generating quantum circuits that can be used as ansatzes in variational algorithms to solve optimization problems.
The agent is trained on diverse problem instances, including Maximum Cut, Maximum Clique and Minimum Vertex Cover, built from different graph topologies and sizes.
Our analysis of the circuits generated by the agent and the corresponding solutions shows that the proposed method is able to generate effective ansatzes.
While our goal is not to propose any new specific ansatz, we observe how the agent has discovered a novel family of ansatzes effective for Maximum Cut problems, which we call $R_{yz}$-connected.
We study the characteristics of one of these ansatzes by comparing it against state-of-the-art quantum algorithms across instances of varying graph topologies, sizes, and  problem types. Our results indicate that the $R_{yz}$-connected circuit achieves high approximation ratios for Maximum Cut problems, further validating our proposed agent.
In conclusion, our study highlights the potential of Reinforcement Learning techniques in assisting researchers to design effective quantum circuits which could have applications in a wide number of tasks.
\end{abstract}

\begin{IEEEkeywords}
Quantum Computing, NISQ, Reinforcement Learning, Variational Quantum Algorithms
\end{IEEEkeywords}

\section{Introduction}
\label{sec:introduction}

Quantum computing has gained attention for its potential to tackle problems that are computationally challenging for classical computers.
Variational Quantum Algorithms (VQAs) \cite{cerezo_2021}, which employ a parametric quantum circuit and a classical optimizer to adjust the circuit parameters, represent a promising paradigm in the Noisy Intermediate-Scale Quantum (NISQ) \cite{preskill_2018} era.
However, a key challenge in these algorithms lies in identifying an appropriate ansatz for the specific problem being addressed.
Approaches for selecting suitable ansatzes include leveraging problem-specific properties, such as symmetries \cite{meyer_2023, le_2023, wierichs_2023}, or using adaptive methods that dynamically modify circuits by adding and removing gates during execution \cite{turati_2023, claudino_2020, mukherjee_2023}.
However, identifying which problem properties can be exploited is non-trivial and adaptive methods rely on carefully designed heuristics and may require numerous circuit executions to converge to a suitable ansatz.

Reinforcement Learning (RL) is a Machine Learning paradigm where an agent learns to perform actions to achieve a desired goal.
By receiving a reward for each action taken, the agent iteratively refines its strategy to maximize the cumulative reward, ultimately achieving the desired objective.
The RL paradigm is highly flexible and has already been successfully deployed for many tasks in quantum computing \cite{wauters_2020, khairy_2020, yao_2020}, including circuit design \cite{pirhooshyaran_2021, ostaszewski_2021}.
One crucial difference between the RL paradigm and the existing adaptive approaches is that it is possible with RL to learn how to build circuits without relying on manually crafted heuristics or domain-specific knowledge. Furthermore, RL can be effective in scenarios with very large solution spaces, such as the one of possible quantum circuits.

In our work we present a RL-based algorithm designed to search for variational quantum circuits for solving optimization problems.
We train the agent on instances of the Maximum Cut, Maximum Clique, and Minimum Vertex Cover problems and analyze the solution quality and the properties of the circuits obtained.
Our findings show that the agent is capable of building circuits leading to satisfactory results, particularly on the Maximum Cut problem.

Moreover, during training on the Maximum Cut problem, the agent discovered effective circuits with a regular structure, which we denote as $R_{yz}$-connected, characterized by circuits with only $R_{yz}$ gates connecting all the qubits.
We test a specific member of this family, referred to as Linear circuit and discuss how they could be implemented on the hardware. Specifically, we test its performance against state-of-the-art quantum algorithms on diverse Quadratic Unconstrained Binary Optimization (QUBO) problems with varying underlining graph topologies and sizes.
Our findings reveal that the Linear circuit is able to find high-quality solutions on the Maximum Cut problem, but it is less effective on the others.

In summary, our contributions are the following:
\begin{itemize}
    \item We propose a Reinforcement Learning agent whose purpose is to generate Variational Quantum Circuits able to solve optimization problems;
    \item We show that the proposed agent is able to build circuits that exhibit good approximation ratios for various optimization problems;
    \item We investigate a specific family of circuits that the agent has discovered, which we call $R_{yz}$-connected, and show that they generalize well and can be used to tackle other instances of the same problem. 
\end{itemize}

Overall, our study shows the potential of RL-based approaches for constructing effective variational quantum circuits. 
Moreover, this methodology holds promise for broader applications within quantum computing, including designing more general circuits or optimizing circuit properties.

\section{Variational Quantum Algorithms}
\label{sec:vqas}

Variational Quantum Algorithms (VQAs) \cite{cerezo_2021} represent a class of hybrid quantum algorithms that can be used to tackle certain optimization problems and are characterized by the presence of a parametric quantum circuit, denoted as ansatz, and a classical optimizer.
The goal of the classical optimizer is to adjust the circuit parameters by minimizing a predefined cost function, in such a way that the execution of the circuit with optimized parameters yields the solution to the given optimization problem.
In recent years, VQAs have known large diffusion due to their potential to tackle complex problems using near-term quantum devices.
Indeed, the variational approach enables the utilization of shallower circuits, making these algorithms more resilient to noise and qubit decoherence.

One of the most widely utilized VQAs is the Variational Quantum Eigensolver (VQE) \cite{peruzzo_2014, tilly_2021}, which allows to determine the ground state of a given Hamiltonian $H$, and is largely applied in quantum chemistry \cite{cao_2019, mcclean_2016}.
Specifically, this algorithm employs a suitable ansatz capable of generating the final parametric state $|\psi(\theta)\rangle$, where $\theta$ denotes a parameter vector.
The objective is to identify the circuit parameters that minimize the cost function $\langle\psi(\theta)|H|\psi(\theta)\rangle$, also denoted as $\langle H\rangle$, through a variational approach.
Notice that, when using real quantum hardware, we do not have direct access to the expectation $\langle\psi(\theta)|H|\psi(\theta)\rangle$, but this quantity can be estimated by executing the circuit a number $n_{\text{shots}}$ of times and computing the quantity:
\begin{equation}
    \label{eq:expectation_estimate}
    \langle H\rangle^* = \langle\psi(\theta)|H|\psi(\theta)\rangle^* = \frac{1}{n_{\text{shots}}} \sum_{i=1}^{n_{\text{shots}}}{\langle\tilde{\psi}_i|H|\tilde{\psi}_i\rangle},
\end{equation}
where $|\tilde{\psi}_i \rangle$ denotes the basis state obtained after performing the measurement at the $i$-th execution of the circuit.

Another widely adopted VQA is the Quantum Approximate Optimization Algorithm (QAOA) \cite{farhi_2014, blekos_2023}, often employed to address combinatorial optimization problems \cite{crooks_2018, willsch_2020, cook_2020, lin_2016, radzihovsky_2019, brandhofer_2022, kurowski_2023}.
Similarly to VQE, QAOA aims to minimize the expectation value of a designated cost Hamiltonian on the circuit's final state.
However, QAOA employs a predefined circuit comprising an initial layer of Hadamard gates followed by $p$ layers, each implementing two parametric operators: the cost operator, which depends on the cost Hamiltonian, and the mixer operator, which allows to further explore the solution space.
QAOA can be considered as a discretized form of continuous-time quantum evolution.

One variant of QAOA is the multi-angle QAOA (ma-QAOA) \cite{herrman_2022}, which shares the same objective and circuit structure as QAOA, but uses distinct parameters for each parametric gate associated with the cost and mixer operators.
Whereas ma-QAOA enhances space exploration capability using the same circuit structure and depth as QAOA, it also increases the computational load on the classical optimizer due to the higher number of parameters, rendering a careful trade-off necessary.

Another variant, QAOA+ \cite{chalupnik_2022}, extends the ansatz used by QAOA with $p=1$ by introducing an additional problem-independent layer consisting of $R_{zz}$ gates, and a mixer layer with $R_x$ gates.
This results in a circuit with Hadamard, cost and mixer layer, and the two newly introduced components.
Notably, this structure is not repeated.
Since each gate in the additional layers is independently parameterized, the QAOA+ ansatz results in a deeper circuit with $2n-1$ additional parameters compared to QAOA with $p=1$.
Similar to ma-QAOA, the higher number of parameters enables a broader exploration of the solution space, but at the same time increases the computational burden on the classical optimizer.

However, one critical challenge in the application of VQAs is the identification of suitable circuits for addressing specific problems \cite{sim_2019, qin_2023, wurtz_2021, du_2020}. 
An ideal ansatz should have a limited number of gates and depth to mitigate noise and decoherence, utilize the native gates of the hardware to simplify implementation and reduce circuit depth, as well as effectively sample the correct solution to the optimization problem.
In particular, one significant obstacle to finding the correct solution to the optimization problem lies in the phenomenon of barren plateaus \cite{mcclean_2018, arrasmith_2021, arrasmith_2022, cerezo_2021_bp, holmes_2022, larocca_2022, volkoff_2021}.
Barren plateaus occur when the gradient of the cost function exponentially vanishes as the system size increases, resulting in a flat landscape in which the classical optimizer struggles to escape local optima and may not be able to find good parameters for the circuit.  
To address this challenge, strategies for selecting suitable ansatzes have been explored.
These include leveraging problem-specific properties\cite{meyer_2023, le_2023, wierichs_2023, farhi_2014} and employing adaptive algorithms\cite{turati_2023, claudino_2020, mukherjee_2023}.

Adaptive Variational Quantum Algorithms are a family of methods which dynamically construct the quantum circuit by iteratively adding and removing gates throughout the algorithm execution.
This approach allows to explore various gate configurations and choose the most suitable circuit structures for a given problem.
One category of adaptive VQAs, proposed in the literature, includes ADAPT-VQE \cite{grimsley_2019}, qubit-ADAPT-VQE \cite{tang_2021}, QEB-ADAPT-VQE \cite{yordanov_2021}, and Overlap-ADAPT-VQE \cite{feniou_2023}, adaptive variants of VQE. These algorithms aim to find the ground state of a Hamiltonian operator representing the energy of a molecule, and their distinguishing feature lies in the construction of the ansatz, which is achieved by adding gates selected from a pool which depends on the specific chemistry problem being addressed.
Additionally, there exists an adaptive version of QAOA \cite{zhu_2022}, which, at each iteration, determines the most appropriate mixer for the following layer of the circuit.
Other adaptive VQAs adopt a genetic approach \cite{rattew_2020, chivilikhin_2020, las-heras_2016}, employ more generalized Machine Learning techniques for circuit construction \cite{cincio_2018, du_2022, bilkis_2023} or make use of Reinforcement Learning as described in \secref{subsec:RL_for_QC}.

\section{Reinforcement Learning}
\label{sec:rl}

Our objective is to use a Machine Learning model to design new circuits capable of finding the ground state of a Hamiltonian. This problem requires exploring a large solution space of potential circuits, and is strongly influenced by the specific Hamiltonian under consideration.
Consequently, creating an exhaustive dataset of quantum circuits for \emph{supervised} learning is impractical. 
Thus, we choose to use a Reinforcement Learning (RL) approach, where an agent interacts with an environment to achieve a specific goal. This enables us to generate data, \ie quantum circuits, and evaluate their performance during training.
Because of the complexity of this topic, in this section we only describe the fundamental concepts of RL and the selected algorithm.
For a comprehensive overview of this technology, the interested reader may refer to \cite{sutton_1998}.

In RL, the agent interacts with the environment in discrete time steps.
At each step $t$, the environment is in a particular \emph{state} $s_t$, which is observed by the agent. Based on this observation, the agent performs an \emph{action} $a_t$, which may have an effect on the environment, causing the transition to a new state $s_{t+1}$. The agent receives a \emph{reward} $r_t$, a value depending on the quality of the action taken.
Starting from an initial state and cyclically interacting with the environment until a termination condition is reached, the agent completes an \emph{episode}.
Let us consider a practical example, where the agent is a robot whose goal is to move an object from one place to another in an environment constituted by a table.
In this scenario, the actions correspond to the movements that the robot's mechanical arms can perform, the state of the environment is represented by the current position of the object on the table, and the reward is a function which increases as the distance between the object and its target location decreases.
An episode starts with the object in an initial position on the table and ends when the robot has successfully placed the object at the target position.

Spanning through multiple episodes, the objective of the agent is to learn which actions to perform in order to obtain the highest cumulative reward, or \emph{return}.
Specifically, the return at time step $t$ is the weighted sum of rewards starting from $t$ until the end of the episode:
\begin{equation}
\label{eq:return}
    g_t = \sum_{k = 0}^{t_e} \gamma^k r_{t+k+1}
\end{equation}
where $t_e$ is the number of remaining steps until the end of the episode and $\gamma \in (0,1)$ represents the discount factor. This factor serves two purposes: facilitating convergence and balancing short-term versus long-term rewards.

When choosing a new action in each state, the agent follows a \emph{policy} $\pi$, which is a probability distribution on the set of all actions, conditioned on the state:
\begin{equation}
    \pi (a|s) = P(a_t = a\ |\ s_t = s)
\end{equation}
Various methods exist to learn an optimal policy.
In this work we use \emph{Proximal Policy Optimization} (PPO), which is currently a state-of-the-art Deep Reinforcement Learning algorithm introduced by OpenAI \cite{ppo_2017}.
PPO makes use of two neural networks with the same structure, respectively called \emph{policy} and \emph{value network}.
Both networks receive an appropriate representation of the environment's state as input.
The policy network then outputs a probability distribution on the possible actions at time step $t$, from which action $a_t$ is then sampled.
On the other hand, the value network outputs a single value estimating the \emph{value function}, which quantifies the quality of the state in terms of expected return when following the policy $\pi$:
\begin{equation}
    \label{eq:value_function}
    V_\pi(s) = \mathbbm{E}_\pi[\ g_t\ |\ s_t = s\ ]
\end{equation}
While the policy network effectively chooses the actions to perform, the value network is used to estimate the so-called \emph{advantage function}:
\begin{equation}
    A_\pi(s, a) = Q_\pi(s, a) - V_\pi(s)
\end{equation}
Here, $Q_\pi$ represents the \emph{state-action value function}, which is the expected return obtained by starting in state $s$, taking action $a$ and then following the policy $\pi$:
\begin{equation}
    Q_\pi(s, a) = \mathbbm{E}_\pi[\ g_t\ |\ s_t = s,\ a_t = a\ ]
\end{equation}
Thus, the advantage measures the potential benefit of taking action $a$ in state $s$, independently of the quality of that state.
In order to learn an optimal policy, PPO performs gradient ascent with the goal of maximizing an objective function that incorporates both the advantage and a loss that helps the value network in learning how to better estimate the real value function.
Moreover, a clipping mechanism is implemented to avoid a too large update to the policy during training.
For further details on the loss and implementation, we refer to the original articles \cite{ppo_2017, spinup_2018}.

\subsection{Reinforcement Learning for Quantum Computing}
\label{subsec:RL_for_QC}

In recent years, RL techniques have been applied to various challenging tasks in quantum computing.
Some studies focus on optimizing circuit parameters \cite{wauters_2020, khairy_2020, yao_2020}, while others tackle circuit learning tasks consisting in generating quantum circuits to transform an initial state into a target state \cite{giordano_2022, kuo_2021, zhu_2023, moro_2021, zhang_2020}.
This task is particularly useful because it can be applied to the crucial step of quantum circuit compiling.
Additionally, RL has been used to learn optimization strategies aimed at reducing circuit depth and gate count \cite{fosel_2021}.
Furthermore, RL algorithms have been applied to the task of designing quantum circuits tailored for Machine Learning \cite{pirhooshyaran_2021} and optimization \cite{ostaszewski_2021} problems.
Specifically, \cite{pirhooshyaran_2021} discusses the use of RL in designing parameterized quantum circuits for classification tasks, while \cite{ostaszewski_2021} employs RL to find suitable ansatzes for VQE to determine the ground state energy of specific molecules.
It is worth noting that the approach taken by \cite{ostaszewski_2021} tackles our same task of finding circuits to generate the ground state of a Hamiltonian.
However, the RL agent is specifically tailored for chemistry problems, and the architecture may not be directly applicable to other domains.
In contrast, our work introduces a novel RL algorithm designed for more general optimization problems and differs in many crucial components: architecture for the agent, state representations, rewards, and available actions.

\section{Agent Design and Training}
\label{sec:agent}

In this study we propose a RL-based algorithm called \emph{Reinforcement Learning for Variational Quantum Circuits} (RLVQC), designed to learn how  to build new quantum circuits aimed at finding the ground state of the Hamiltonian associated with a given optimization problem.
In this section we describe the different components of RLVQC, including environment, actions, and reward.
Then, we present the experimental setup and specify some implementation details useful to reproduce the experiments.

\subsection{Environment, Actions, Reward}
\label{subsec:env_actions_reward}
In RLVQC, the environment is represented by a parametric circuit with $n$ qubits. At each step $t$, the agent performs an action consisting in adding a new gate to the circuit.
The environment configuration at the beginning of each episode is represented by a circuit with a single layer of Hadamard gates.
The action set $\mathcal{A}$ comprises the gates that the agent can insert into the circuit. Specifically, $\mathcal{A}$ is the union of the following sets:
\begin{itemize}
    \item $\mathcal{S} = \{R_a^i(\theta) \ | \ a \in \{x, y, z\}, \ i = 0,..., n - 1\}$
    \item $\mathcal{D} = \{R_{ab}^{ij}(\theta) \ | \  a,b \in \{x, y, z\}, \ i,j = 0,...,n-1, \ i < j\}$
\end{itemize}
where $R_a^i$ is a single rotation gate applied to qubit $i$ and $R_{ab}^{ij}$ is a double rotation applied to a pair of different qubits $i$ and $j$.
Formally, denoting as $\sigma_a$ the Pauli-$a$ matrix for all $a,b\in\{x,y,z\}$, the double rotations are defined as:
\begin{equation}
    \label{eq:double_rotations}
    R_{ab}(\theta) = e^{-i\frac{\theta}{2}\sigma_a\otimes\sigma_b}
\end{equation}
A specific double rotation $R_{ab}(\theta)$ can be decomposed in simpler gates as follows:
\begin{equation*}
    \vcenter{\hbox{
        \begin{quantikz}
        & \gate[2]{R_{ab}(\theta)} &  \qw \\
        & \qw & \qw
        \end{quantikz}
    }}
    \ = \
    \vcenter{\hbox{
        \begin{quantikz}
        & \gate{U_b} & \gate[2]{R_{zz}(\theta)} & \gate{U_b^\dag} &  \qw \\
        & \gate{U_a} & \qw & \gate{U_a^\dag} & \qw
        \end{quantikz}
    }}
\end{equation*}
where $U_a$ and $U_b$ are gates which map the $a$ and $b$ axis into the $z$ axis respectively. Thus, $U_x = H$, $U_y = R_x(\frac{\pi}{2})$, and $U_z = I$.
These double rotations are able to generate entanglement between qubits.

The gates in the set $\mathcal{A}$ are chosen because they exhibit identity-like behavior when their parameters are set to 0.
This property is advantageous because starting from a circuit with optimized parameters and adding a new gate with parameters set to 0 allows the optimization process to begin from a favorable initial point rather than a random one.
This approach potentially reduces computational time and mitigates the risk of converging to a local minimum\cite{grant_2019}.

At each time step $t$, the agent chooses a gate as an action to be applied to the environment, which is updated by adding that gate with parameter $\theta_t = 0$ to the circuit.
Notice that all the gates are independently parametrized.
Subsequently, the circuit parameters are optimized using the classical optimizer COBYLA\footnote{We use the SciPy implementation of COBYLA, keeping its default hyperparameters. The documentation is available here: \url{https://docs.scipy.org/doc/scipy/reference/optimize.minimize-cobyla.html}}, which has been shown to be effective in noise-free scenarios without demanding an excessive computational cost \cite{singh_2023, fernandez-pendas_2020}.
COBYLA is run for a maximum of 1000 iterations with the goal of minimizing the cost function given by the expectation of the problem Hamiltonian on the final state of the circuit, defined in \eqref{eq:expectation_estimate}.
Such expectation is estimated by simulating the circuit 1000 times\footnote{Circuit simulation is performed using Qiskit QASM Simulator. The documentation is available here: \url{https://qiskit.github.io/qiskit-aer/stubs/qiskit_aer.QasmSimulator.html}}.
The choice of estimating the expectation instead of computing it exactly is made to reflect the use of an actual quantum computer.
Indeed, on real quantum devices, we do not have direct access to the quantum state, but we can only estimate it by performing multiple measurements.

After COBYLA converges, the circuit using the optimized parameters is executed an additional 1000 times to obtain the estimated probability distribution of the final state of the circuit, in the form of a vector of $2^n$ real-valued elements.
This probability distribution represents the next state observed by the agent.
Notice that the design of an effective representation for the environment state is not trivial and depends on several factors related to the specific task and the agent's architecture.
For this reason, designing new state representations specifically tailored for quantum systems can be considered an important and wide research area that goes beyond the scope of this paper.

The reward at time step $t$ is defined to align with our goal of minimizing the expectation value of the Hamiltonian and the circuit depth while the agent learns to maximize the return \eqref{eq:return}. Specifically, the reward is expressed as:
\begin{equation}
\label{eq:reward}
    r_t = -\langle H\rangle^*_t - \beta\cdot d_t,
\end{equation}
where $\langle H\rangle^*_t$ is an estimate of the current expectation value as defined in \eqref{eq:expectation_estimate} and $d_t$ denotes the current circuit depth, weighted by the $\beta$ hyperparameter, which is fixed at $0.015$ in our experiments.
This reward is straightforward and directly tied to the quality of the circuit built at time step $t$: the first term incentivizes the minimization of $\langle H\rangle^*$, which is our primary goal, while the second term drives the agent to prefer shallower circuits, which are more resilient to noise and qubit decoherence.
It is worth noting that in this phase the circuit depth is computed after expressing the circuits in terms of the basis gates $\{H, R_x, R_y, R_z, R_{zz}\}$.
In particular, while the double rotation $R_{zz}$ is directly used, the others are represented in terms of it. 
Finally, we remark that designing an effective reward function that combines different goals is again a critical aspect in RL which allows for great flexibility.
Moreover, it is important to highlight that the suitability of a particular reward function may vary depending on the nature of the task and specific requirements.

\begin{figure}
  \centering
  \begin{subfigure}[t]{\columnwidth}
        \centering
        \begin{tikzpicture}[
            box/.style={rectangle, draw=black, thick},
            ]
            \node[box](agent){Agent};
            \node[box](env)[right=of agent]{Environment};
            
            \draw[dashed,->] ([yshift=-0.92cm]$([xshift=5]env.south)!0.8!([xshift=-7]agent.south)$) .. controls +(left:0) and +(down:0.75) .. ([xshift=-7]agent.south)
                node[pos=0.7,left]{$s_t$};
            
            \draw[->] (agent.north) .. controls +(up:1.2) and +(up:1.2) .. (env.north)
                node[pos=0.5,below]{$a_t$};
            
            \draw[->] (env.south) .. controls +(down:1.2) and +(down:1.2) .. (agent.south)
                node[pos=0.5,above]{$s_{t+1}$}
                node[pos=0.5,below]{$r_t$};
        \end{tikzpicture}
        \caption{Interaction between agent and environment. At time step $t$ the agent observes the environment's state $s_t$ and acts with action $a_t$ on the environment, which gives reward $r_t$ to the agent. At the next time step, the agent will observe the new state $s_{t+1}$.}
        \label{fig:rl:int}
    \end{subfigure}
    \par\medskip
    \begin{subfigure}[t]{\columnwidth}
        \centering
        \begin{tikzpicture}[
                box/.style={minimum width=4mm, minimum height=4mm},
                cell/.style={minimum width=3mm, minimum height=3mm},
                network/.style={minimum width=20mm, minimum height=8mm},
                line/.style={solid},
                dline/.style={densely dashed},
            ]
    
            \coordinate (base) at (-20mm,-14mm);
            \coordinate (actions) at (30mm,-14mm);
            \coordinate (action) at (42mm,-6mm);
            \coordinate (fillpad) at ($(base) + (4mm,0)$);
            \coordinate (actionpad) at ($(action) + (-4mm,0)$);
    
            \node[network,draw] (pn) at (0,0) {Policy Net};
            \node[network,draw,below=4mm of pn] (vn) {Value Net};
    
            \node[fit=(pn)(vn)(fillpad)(actions)(actionpad),inner sep=4mm,draw] (agent) {};
            \node[above=0mm of agent]{Agent};
    
            \foreach [count=\i from 0,evaluate=\i as \y using \i*4] \shade in {90, 50, 30, 0, 20}{
                \node [box,draw,fill=white] (s\i) at ($(0,\y mm) + (base)$) {};
                \node [cell,draw,fill=Firebrick1!\shade!white,rounded corners=0.8mm] (c\i) at ($(0,\y mm) + (base)$) {};
            }
            \node[box,draw,fill=white,text width=2.7mm,inner sep=0pt] (etc) at ($(0,8mm) + (base)$) {...};
            \draw[white, very thick, dotted] ($(-2mm,6.4mm) + (base)$) -- ($(-2mm,9.6mm) + (base)$) ($(2mm,6.4mm) + (base)$) -- ($(2mm,9.6mm) + (base)$);
            
            \draw[->] ([xshift=-8mm]etc.west) -- (etc.west)
                node[near start,above]{$s_t$};
    
            \draw[->] (etc.east) .. controls ($(4mm,0) + (etc.east)$) and ($(-4mm,0) + (pn.west)$) .. (pn.west) {};
            \draw[->] (etc.east) .. controls ($(4mm,0) + (etc.east)$) and ($(-4mm,0) + (vn.west)$) .. (vn.west) {};
    
            \foreach [count=\i from 0,evaluate=\i as \y using \i*4] \shade in {10, 60, 30, 120, 30}{
                \node [box,draw,fill=white] (a\i) at ($(0,\y mm) + (actions)$) {};
                \node [cell,draw,fill=RoyalBlue3!\shade!white,rounded corners=0.8mm] (ca\i) at ($(0,\y mm) + (actions)$) {};
            }
            \node[box,draw,fill=white,text width=2.7mm,inner sep=0pt] at ($(0,8mm) + (actions)$) {...};
            \draw[white, very thick, dotted] ($(-2mm,6.4mm) + (actions)$) -- ($(-2mm,9.6mm) + (actions)$) ($(2mm,6.4mm) + (actions)$) -- ($(2mm,9.6mm) + (actions)$);
    
            \draw[->] (pn.east) -- ([xshift=-2mm,yshift=14mm]actions.west) {}
                node[below,midway] {$\pi(a|s_t)$};
    
            \node[right=4mm of vn] (value) {$\hat{V}_\pi(s_t)$};
            \draw[->] (vn.east) -- (value.west) {};
            
            \node[draw,fill=white] (act) at (action) {$a_t$};
            \draw[-] ($(2.8mm,18mm) + (actions)$) .. controls ($(4mm,0) + (a4.north east)$) and ($(-4mm,0) + (act.north west)$) .. ([yshift=-0.07mm]act.north west) {};
            \draw[-] ($(2.8mm,-2mm) + (actions)$) .. controls ($(4mm,0) + (a0.south east)$) and ($(-4mm,0) + (act.south west)$) .. ([yshift=0.07mm]act.south west) {};
            \draw[->] (act.east) -- ([xshift=8mm]act.east);
        \end{tikzpicture}
        \caption{State $s_t$ is processed by the agent's neural networks. The value network outputs an estimate $\hat{V}_\pi(s_t)$ of the value function \eqref{eq:value_function}, while the policy network outputs a probability distribution $\pi(a|s_t)$ on the actions. Action $a_t$ is sampled from this probability distribution.}
        \label{fig:rl:agent}
    \end{subfigure}
    \par\medskip
    \begin{subfigure}[t]{\columnwidth}
        \centering
        \begin{tikzpicture}[
                box/.style={minimum width=4mm, minimum height=4mm},
                cell/.style={minimum width=3mm, minimum height=3mm},
                network/.style={minimum width=20mm, minimum height=8mm},
                line/.style={solid},
                dline/.style={densely dashed},
            ]
            \node[scale=0.8] (circ) {
            \begin{quantikz}[row sep=0.2em, column sep=0.8em]
                \lstick{$q_2 :$} & \gate{\mathrm{H}} & \ \ldots\ & \ctrl{1} & \qw & \ctrl{1} & \qw & \meter{} \\
                \lstick{$q_1 :$} & \gate{\mathrm{H}} & \ \ldots\ & \targ{} & \gate{R_z(\theta_{t-1})} & \targ{} & \qw & \meter{} \\
                \lstick{$q_0 :$} & \gate{\mathrm{H}} & \ \ldots\ & \qw & \qw & \qw & \qw & \meter{}
            \end{quantikz}
            };
            \node[above left=-2mm of circ,anchor=south west] (ts) {Time step $t-1$};
            \node[above=0mm of ts] (fillts) {};
    
            \node[scale=0.8,below=6mm of circ] (postcirc) {
            \begin{quantikz}[row sep=0.2em, column sep=0.8em]
                \lstick{$q_2 :$} & \gate{\mathrm{H}} & \ \ldots\ & \ctrl{1} & \qw & \ctrl{1} & \qw & \qw & \qw & \meter{} \\
                \lstick{$q_1 :$} & \gate{\mathrm{H}} & \ \ldots\ & \targ{} & \gate{R_z(\theta_{t-1})} & \targ{} & \qw & \qw & \qw & \meter{} \\
                \lstick{$q_0 :$} & \gate{\mathrm{H}} & \ \ldots\ & \qw & \qw & \qw & \qw & \gate{R_x(\theta_t)} \gategroup[1,steps=1,style={dashed},label style={label position=below,yshift=-4mm}]{$a_t$} & \qw & \meter{}
            \end{quantikz}
            };
            \node[above left=-2mm of postcirc,anchor=south west]{Time step $t$};
    
            \draw[->] (circ.south) -- ([yshift=-2mm]postcirc.north) {};
    
            \node[box,inner sep=2mm,below=4mm of postcirc,align=center,draw] (opt) {Optimization\\+\\Simulation};
            \draw[->] ([yshift=6mm]postcirc.south) -- ([yshift=2mm]opt.north) {};
    
            \node[fit=(fillts)(circ)(postcirc)(opt),inner sep=4mm,draw] (env) {};
            \node[above=0mm of env]{Environment};
    
            \node[box,draw,fill=white,above left=2mm of fillts] (action) at (fillts) {$R_x(\theta_t)$};
            \draw[->] ([yshift=8mm]action.north) -- (action.north) {}
                node[left,midway] {$a_t$};
    
            \node[draw,box,below right=0mm and 8mm of opt,inner sep=2mm,outer sep=2mm,fill=white,align=center] (state) {$s_{t+1}$\\$r_t$};
            \draw[->] ([xshift=2mm]opt.east) .. controls ($(opt) + (27mm,0)$) and ($(state) + (0,15mm)$) .. (state.north) {};
            \draw[->] ([xshift=-2mm]state.east) -- ([xshift=8mm]state.east) {};
    
        \end{tikzpicture}
        \caption{When the environment receives action $a_t$, the corresponding gate is added to the circuit. Then, its parameters are optimized and the circuit is simulated to obtain the next state $s_{t+1}$, which is sent back to the agent with the corresponding reward $r_t$.}
        \label{fig:rl:env}
    \end{subfigure}

  \caption{Visual overview of the RL pipeline, describing the interaction between agent and environment (see \ref{fig:rl:int}) and how the agent (see \ref{fig:rl:agent}) and environment (see \ref{fig:rl:env}) work internally.}
\label{fig:rl}
\end{figure}
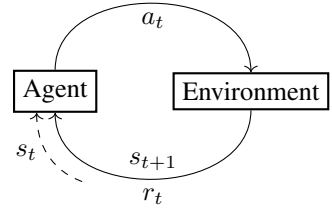
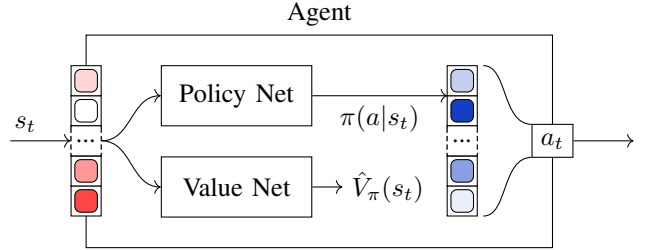
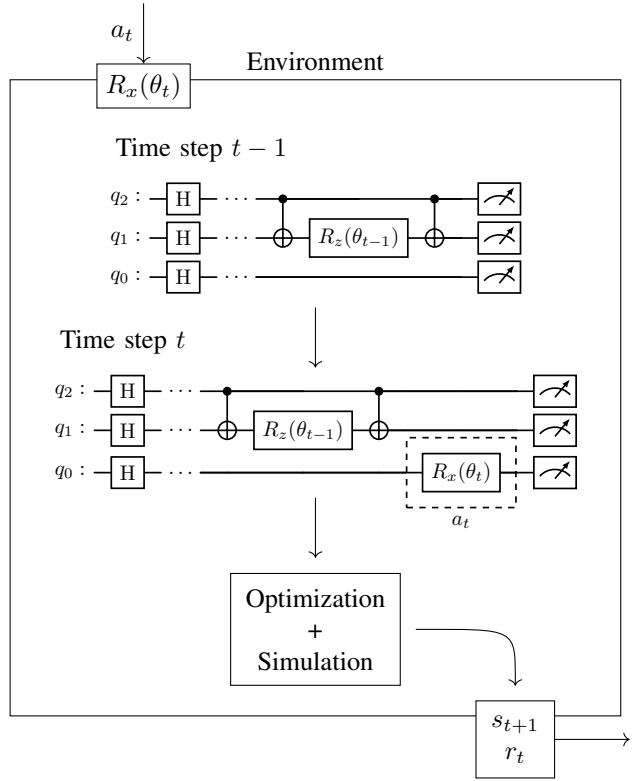

A complete visual overview of the RL pipeline with a detailed description of the agent and environment is represented in \figref{fig:rl}.

\subsection{Training Details}
RLVQC uses the version of PPO implemented as described by OpenAI \cite{ppo_2017, spinup_2018}, with default hyperparameters.
This implementation employs two multi-layer fully-connected neural networks, each sharing the same structure but with separate learnable parameters, for the policy and value network.
With the problem size denoted as $n$, the input layer of each neural network comprises $2^n$ neurons, matching the size of the vector representing the environment state (see Section \ref{subsec:env_actions_reward}).
The output layer for the policy network has size $|\mathcal{A}|$, which depends on the number of problem variables.
Agent training comprises 64 \emph{epochs}, with each epoch being a collection of 384 action steps\footnote{This number is obtained by having multiple processes performing 64 action steps each in our default implementation.} after which the parameters of the PPO's neural networks are updated.
Each epoch consists of multiple episodes, each starting from a circuit with only Hadamard gates and ending when a termination condition is met.
Specifically, an episode terminates either when a maximum of $2\cdot n$ actions have been performed or earlier, if the reward does not improve with new actions.
The latter mechanism is controlled by the \emph{patience} hyperparameter.
Patience is initially set to 3 and decreases every time a step ends with a worse reward than the highest found during the current episode.
Otherwise, it is increased, but only up to its initial value.
Stopping the episode earlier avoids spending time steps on actions that are unlikely to improve the reward.
Throughout training, we keep track of the circuit with the highest reward, which is then analyzed after training is completed.

\subsection{Problem Instances}

In our experiments, we evaluate RLVQC on the task of finding the solution of optimization problems that can be formulated as \emph{Quadratic Unconstrained Binary Optimization} (QUBO) problems \cite{glover_2022}. The general formulation of a QUBO problem is the following:
\begin{equation}
\label{eq:qubo}
    \quad \min_{x\in\{0,1\}^n} x^T \mathbf{Q}  x
\end{equation}
where $x=\{x_1,\dots,x_n\}\in \{0,1\}^n$ and $\mathbf{Q}$ denotes an upper triangular or symmetrical $n \times n$ matrix.
QUBO problems can be expressed as equivalent Ising problems through a change of variable \cite{lucas_2014}.
Leveraging this formulation, the objective becomes to identify the ground state of a Hamiltonian operator, a task which can be accomplished using our algorithm RLVQC.

We address three diverse optimization problem types: Maximum Cut, Maximum Clique, and Minimum Vertex Cover, all formulated as QUBO problems.\cite{glover_2022, pelofske_2019}.
It is important to highlight that Minimum Vertex Cover and Maximum Clique involve constraints. To incorporate these constraints into the QUBO formulation, penalty terms are directly integrated into the cost functions. Each penalty term is a function that yields a value of 1 when a constraint is violated and 0 when it is satisfied, multiplied by a weight coefficient.
These coefficients are chosen to ensure that all the feasible solutions have a lower expectation than the infeasible ones, thereby guiding the algorithm towards favoring feasible solutions over infeasible ones.
For each problem type, we consider three different graph topologies:
\begin{itemize}
    \item 3-regular, a graph where each vertex is connected to exactly three neighbors;
    \item 2D-grid, a two-dimensional integer lattice graph where each vertex can have up to 4 neighbours;
    \item Star graph, characterized by one central node connected to all other nodes, with no further connections among the non-central nodes.
\end{itemize}
For each problem type, we explore graphs with $n=8$ and $n=14$ vertices across each topology, resulting in a total of 18 trained agents.

\subsection{Evaluation Metrics}
\label{subsec:metrics}

Our main goal is to assess whether it is possible to use a RL agent to discover ansatzes capable of sampling high-quality solutions to optimization problems.
We aim to evaluate both the accuracy of the solutions found and the characteristics of the resulting circuits.
To this purpose, the most relevant metric we use is the \emph{approximation ratio} (A.R.), defined as:
\begin{equation}
    \text{A.R.} = \frac{\langle H \rangle^* - \langle H \rangle_{\text{max}}}{\langle H \rangle_{\text{min}} - \langle H \rangle_{\text{max}}},
\end{equation}
where $\langle H \rangle_{\text{min}}$ and $\langle H \rangle_{\text{max}}$ represent the minimum and maximum values of the achievable expectations, respectively, and $\langle H \rangle^*$ is an estimate of the expectation, as defined in \eqref{eq:expectation_estimate}.
The intuition behind this metric is that the ratio $\frac{\langle H \rangle^*}{\langle H \rangle_{\text{min}}}$ approaches 1 when $\langle H \rangle^*$ closely approximates $\langle H \rangle_{\text{min}}$, indicating that the circuit allows to sample with high probability states whose energy values are close to the minimum.
To guarantee that both the numerator and the denominator of the ratio have negative values, we subtract $\langle H \rangle_{\text{max}}$ from both quantities.
This ensures that the final approximation ratio falls between 0 and 1, with closer proximity to 1 when the solution quality is higher.
It is worth noting that in the case of the Maximum Cut problem, $\langle H \rangle_{\text{max}} = 0$.

However, the approximation ratio alone does not indicate if the constraints in Minimum Vertex Cover and Maximum Clique are satisfied.
Hence, in \tabref{tab:rl}, we include a threshold corresponding to the lowest approximation ratio of feasible solutions.
Circuits with approximation ratios below this threshold are more likely to sample infeasible solutions.
Notice that, for the unconstrained Maximum Cut problem, the feasibility threshold is 0.

Finally, we analyze the composition of the final circuit, reporting the number of single and two-qubit gates, as well as the circuit depth.
These metrics are computed after expressing the circuit in terms of the gates $\{Cx, R_x, R_y, R_z\}$, where $Cx$ denotes the CNOT gate, and  and applying some simplifications using the Qiskit transpiler with optimization level 1, which is the default setting.

\section{Results and Discussion}
\label{sec:results}

In this section, we discuss the results obtained by executing the RLVQC algorithm according to the experimental protocol outlined in \secref{sec:agent}, using the metrics detailed in \secref{subsec:metrics}.
To establish a baseline for comparison, we also run QAOA with $p=1$ on the same problem instances and analyze the same set of metrics. Since QAOA is a stochastic algorithm which strictly depends on the initial parameters, we perform 10 executions of the algorithm for each problem instance. We choose random initial parameters for each run, and then average the approximation ratios obtained.
This comparative analysis is presented in \secref{subsec:agent_results}.

Notably, during training on the Maximum Cut problem, the agent discovered circuits with a common structure and high approximation ratios.
This observation led us to identify a novel family of ansatzes, which we denote as $R_{yz}$-connected circuits, as discussed in \secref{subsec:ryz_connected}.
In particular, we conduct an analysis of the most intuitive circuit within this family, which we refer to as \emph{Linear} circuit, by examining its performance across a broader set of graph topologies and different problem types.
Finally, in \secref{subsec:implementability}, we discuss the straightforward implementability of the $R_{yz}$-connected circuits.

\subsection{RLVQC Results}
\label{subsec:agent_results}

\begin{table*}
    \centering
    \begin{tabular}{c|l|l|cc|c|cc|cc|cc}
    \toprule
    \multirow{2}{*}{Size} & \multirow{2}{*}{Problem} & \multirow{2}{*}{Graph} & \multicolumn{2}{c|}{A.R.} & \multirow{2}{*}{A.R. Thresh.} & \multicolumn{2}{c|}{Single-qubit} & \multicolumn{2}{c|}{Two-qubit} & \multicolumn{2}{c}{Depth}\\
    &&& RLVQC & QAOA & & RLVQC & QAOA & RLVQC & QAOA & RLVQC & QAOA \\
    \midrule{}
    \multirow{9}{*}{8}
    &\multirow{3}{*}{\shortstack[l]{Maximum \\ Cut}} 
    &3-regular&$0.99$ & $0.75$ & $0.00$ & $37$ & $36$ & $14$ & $24$ & $38$ & $12$\\
    &&2D-grid &$0.99$ & $0.63$ & $0.00$ & $37$ & $34$ & $14$ & $20$ & $38$ & $12$\\
    &&Star graph & $0.99$ & $0.70$ & $0.00$ & $37$ & $31$ & $14$ & $14$ & $38$ &$24$\\ \cmidrule{2-12}
    &\multirow{3}{*}{\shortstack[l]{Maximum \\ Clique}}
    &3-regular &$0.75$ & $0.81$ & $0.91$ & $62$ & $48$ & $22$& $32$ & $40$ & $15$\\
    &&2D-grid &$0.83$ & $0.78$ & $0.94$ & $47$ & $50$ & $20$& $36$ & $43$ & $15$\\
    &&Star graph &$0.70$ & $0.80$ & $0.95$ & $44$ & $53$ & $18$& $42$ & $38$ & $21$\\ \cmidrule{2-12}
    &\multirow{3}{*}{\shortstack[l]{Minimum \\ Vertex \\ Cover}}
    &3-regular &$0.99$ & $0.83$ & $0.77$ & $63$ & $44$ & $18$ & $24$ & $43$ & $12$\\
    &&2D-grid &$0.96$ & $0.85$ & $0.73$ & $28$ & $42$& $6$& $20$& $16$ & $12$\\
    &&Star graph &$0.93$ & $0.83$ & $0.69$ & $31$ & $39$& $8$& $14$& $20$ & $24$\\ \midrule
    \multirow{9}{*}{14}
    &\multirow{3}{*}{\shortstack[l]{Maximum \\ Cut}}
    &3-regular &$0.92$ & $0.74$ & $0.00$ & $67$ & $63$ & $26$& $42$ & $56$ & $12$\\
    &&2D-grid &$0.76$ & $0.65$ & $0.00$ & $64$ & $61$ & $24$& $38$ & $52$ & $12$\\
    &&Star graph &$0.50$ & $0.71$ & $0.00$ & $49$ & $55$ & $14$& $26$& $32$ & $42$\\ \cmidrule{2-12}
    &\multirow{3}{*}{\shortstack[l]{Maximum \\ Clique}}
    &3-regular &$0.64$ & $0.79$ & $0.98$ & $80$ & $126$ & $26$& $140$ & $48$ & $34$\\
    &&2D-grid &$0.72$ & $0.78$ & $0.99$ & $78$ & $128$& $24$& $144$ & $35$ & $34$\\
    &&Star graph &$0.63$ & $0.76$ & $0.99$ & $97$ & $134$& $34$& $156$& $55$ & $39$\\ \cmidrule{2-12}
    &\multirow{3}{*}{\shortstack[l]{Minimum \\ Vertex \\ Cover}}
    &3-regular &$0.85$ & $0.78$ & $0.74$ & $87$ & $77$ & $28$& $42$& $54$ & $12$\\
    &&2D-grid &$0.87$ & $0.87$ & $0.74$ & $72$ &$75$ & $18$& $38$& $33$ & $12$\\
    &&Star graph &$0.65$ & $0.82$ & $0.71$ & $55$ & $69$ & $14$& $26$ & $32$ & $42$\\ \bottomrule
    \end{tabular}
    \caption{Comparative analysis of the performance of RLVQC and QAOA with $p=1$ on instances of different problem types built from graph with various topologies and $n=8$ and $n=14$ vertices. 
    The table reports the approximation ratios relative to the circuits with the highest reward found by RLVQC and the average approximation ratio reached by QAOA across 10 executions of the algorithm, along with the threshold for feasibility. 
    Additionally, the table shows the number of single-qubit and two-qubit gates, as well as the circuit depth.}
    \label{tab:rl}
\end{table*}

In \tabref{tab:rl} we report the results relative to the solutions and circuits found with RLVQC and QAOA with $p=1$ in terms of the metrics described in \secref{subsec:metrics}.
Specifically, the table shows the approximation ratio (A.R.) of the solutions, the approximation ratio threshold (A.R. Thresh.) for feasibility, the number of single and two-qubit gates, and the circuit depth.

By examining \tabref{tab:rl}, it can be noticed that the approximation ratios found by RLVQC is significantly influenced by the type of optimization problem being addressed.
RLVQC consistently outperforms QAOA in Maximum Cut and Minimum Vertex Cover instances, with one exception observed in the case of the star-topology graph instances with $n=14$ vertices for both problem types, where RLVQC performs worse.
Moreover, for the Minimum Vertex Cover instance with star-topology and $n=14$, RLVQC falls below the threshold for feasibility in terms of approximation ratio.
The particularly strong performance of RLVQC on Maximum Cut instances, especially on instances with $n=8$ qubits where RLVQC consistently achieves an approximation ratio of 0.99, prompted us to conduct a deeper analysis of these circuits.
Interestingly, we observed that the agent exclusively added $R_{yz}$ rotations to these circuits during their construction, without performing any single-qubit actions.
This observation suggests that circuits containing $R_{yz}$ rotations may be particularly effective for addressing Maximum Cut problems.
In contrast, for Maximum Clique instances, RLVQC consistently yields lower approximation ratios compared to QAOA, except for the instance built from a 2D-grid-topology graph with $n=8$ vertices.
However, neither algorithm attains satisfactory results, as the threshold for feasibility is never met for any graph topology.

Upon analyzing the count of single and two-qubit gates, as well as the circuit depth, we observe that RLVQC generally employs a number of gates comparable to QAOA but exhibits a higher depth.
Notably, there are cases where the RLVQC circuit has a gate count that is significantly lower than in QAOA, yet the depth increases.
An example is given by Maximum Clique with $n=14$, where the count of two-qubit gates amounts to less than 25\% of the two-qubit gates used by QAOA.
Finally, it is important to notice that, while circuits produced by RLVQC tend to have higher depth compared to QAOA, adjusting the $\beta$ hyperparameter in the reward function \eqref{eq:reward} allows for further penalizing deep circuits.
This adjustment may guide RLVQC toward the preference of low-depth circuits.

In summary, RLVQC successfully found solutions with approximation ratios comparable to QAOA, or superior in the case of the Maximum Cut problem.
It is noteworthy that these results are obtained using a relatively straightforward design for the RL agent, environment and reward.
Therefore, there is a considerable potential to further improve these results by better understanding how to design these components for quantum computing tasks.

\subsection{$R_{yz}$-connected Ansatzes}
\label{subsec:ryz_connected}
Here we describe the family of $R_{yz}$-connected ansatzes discovered during training on the Maximum Cut problem.
This family is composed of circuits with an initial layer of Hadamard gates, followed by $n - 1$ $R_{yz}$ rotations, which are applied such that each new $R_{yz}$ adds only one qubit to the ``chain'' of connected qubits. More formally, calling $C_j$ the graph representing the connectivity of the circuit after applying the first $j$ $R_{yz}$ rotations, the $C_j$ graphs associated to each $R_{yz}$-connected ansatz are connected for all $j=1,\dots,n-1$.

Recall that a $R_{zz}$ gate is equivalent to two $Cx$ gates with a $R_z$ rotation between them on the target qubit, as shown below:
\begin{equation*}
    \vcenter{\hbox{
        \begin{quantikz}
        & \gate[2]{R_{zz}(\theta)} & \qw \\
        & \qw & \qw
        \end{quantikz}
    }}
    \ = \
    \vcenter{\hbox{
        \begin{quantikz}
            & \ctrl{1} & \qw & \ctrl{1} & \qw \\
            & \targ{} & \gate{R_z(\theta)} & \targ{} & \qw
        \end{quantikz}
    }}
\end{equation*}
Using the more general definition of double rotations provided in \eqref{eq:double_rotations} and the above expansion of the $R_{zz}$ gate, we have that $R_{yz}$ rotations can be decomposed as follows:
\begin{equation*} \label{fig:ryz}
\resizebox{\columnwidth}{!}{
    $\vcenter{\hbox{
        \begin{quantikz} 
        & \gate[2]{R_{yz}(\theta)} & \qw\\
        & \qw & \qw
    \end{quantikz}
    }}
    \ = \
    \vcenter{\hbox{
        \begin{quantikz}
        & \qw & \ctrl{1} & \qw & \ctrl{1} & \qw & \qw \\
        & \gate{R_x(\frac{\pi}{2})} & \targ{} & \gate{R_z(\theta)} & \targ{} & \gate{R_x(-\frac{\pi}{2})} & \qw        
        \end{quantikz}
    }}$
}
\end{equation*}

A notable property of $R_{yz}$-connected circuits is that two basis states that have all bits flipped \wrt to one another have the same probability of being measured.
This characteristic makes the circuits especially well-suited for problems that possess this symmetry property, such as the Maximum Cut problem.

\begin{figure*}[ht]
    \centering
\resizebox{\textwidth}{!}{
\begin{quantikz}[row sep=0.2em, column sep=0.8em]
    \lstick{$q_3 :$} & \gate{\mathrm{H}} & \qw & \ctrl{1}
    \gategroup[2,steps=4,style={dashed,inner xsep=0.33em, inner ysep=0.33em}]{}
    & \qw & \ctrl{1} & \qw & \qw & \qw & \qw & \qw & \qw & \qw & \qw & \qw & \qw & \qw & \qw & \meter{}\\
    \lstick{$q_2 :$} & \gate{\mathrm{H}} & \qw & \targ{} & \gate{R_z(\theta_{23})} & \targ{} & \gate{R_x(-\frac{\pi}{2})} & \qw & \ctrl{1}
    \gategroup[2,steps=4,style={dashed,inner xsep=0.33em, inner ysep=0.33em}]{}
    & \qw & \ctrl{1} & \qw & \qw & \qw & \qw & \qw & \qw & \qw & \meter{}\\
    \lstick{$q_1 :$} & \gate{\mathrm{H}} & \qw & \qw & \qw & \qw & \qw & \qw & \targ{} & \gate{R_z(\theta_{12})} & \targ{} & \gate{R_x(-\frac{\pi}{2})} & \qw & \ctrl{1}
    \gategroup[2,steps=4,style={dashed,inner xsep=0.33em, inner ysep=0.33em}]{}
    & \qw & \ctrl{1} & \qw & \qw & \meter{}\\
    \lstick{$q_0 :$} & \gate{\mathrm{H}} & \qw & \qw & \qw & \qw & \qw & \qw &\qw & \qw & \qw & \qw & \qw & \targ{} & \gate{R_z(\theta_{01})} & \targ{} & \gate{R_x(-\frac{\pi}{2})} & \qw & \meter{}
\end{quantikz}
}
\caption{Linear circuit, a specific element of the family of $R_{yz}$-connected ansatzes found during training on Maximum Cut. \label{c:linear}}
\end{figure*}
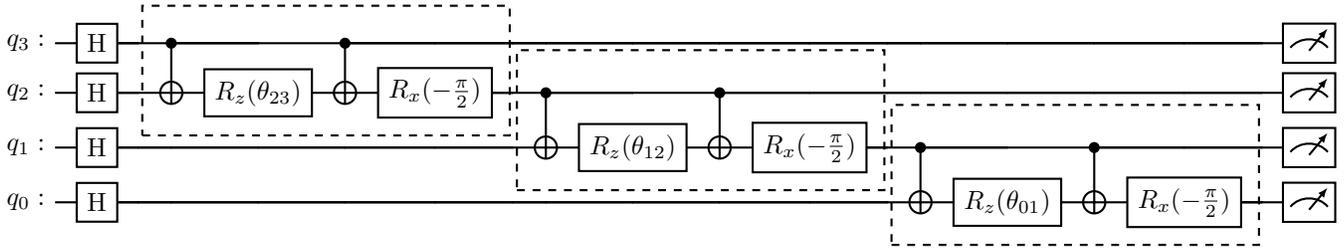

Among the elements of the $R_{yz}$-connected family, we focus on a specific circuit, which we denote as \emph{Linear} circuit, where each $R_{yz}$ gate connects one qubit to the next one (see \figref{c:linear}). It is worth noting that each block of the figure represents the decomposition of a $R_{yz}$ gate as explained in \secref{fig:ryz}.
The first $R_x(\frac{\pi}{2})$ gate is omitted since it behaves as the identity when applied after a Hadamard gate.
The choice of analyzing this particular $R_{yz}$-connected ansatz is motivated by its uniform application of $R_{yz}$ gates across all qubits, making it the most straightforward configuration within the family.

The primary objective of our analysis is to evaluate the quality of solutions obtained by the Linear circuit on the Maximum Cut problem, but across a broader range of underlying graph topologies.
We maintain the use of 3-regular, 2D-grid, and star graphs, and further include cycle graphs, where each vertex is connected to exactly two other vertices, forming a closed loop. Additionally, we introduce Erdős–Rényi graphs generated from a set of vertices, with edges independently connected based on fixed probabilities of 0.2, 0.5, and 0.8.
Subsequently, we evaluate the performance of the Linear circuit on other problem types, specifically Maximum Clique and Minimum Vertex Cover, using the same underlying topologies.
The algorithms are tested on instances with 8, 14, and 16 vertices, which represents the largest tractable size with our available hardware.
However, we present results only for the 16-vertex instances, as the findings for smaller instances are analogous.
We compare the performance of the Linear circuit with state-of-the-art quantum algorithms, including QAOA with depths $p=1$ and $p=2$, QAOA+, and ma-QAOA (see \secref{sec:vqas}). Each algorithm is executed 10 times on each problem instance and the results are averaged. We choose random initial parameters for each execution.

\begin{figure*}[ht]
    \centering
    \includegraphics[width=1\textwidth]{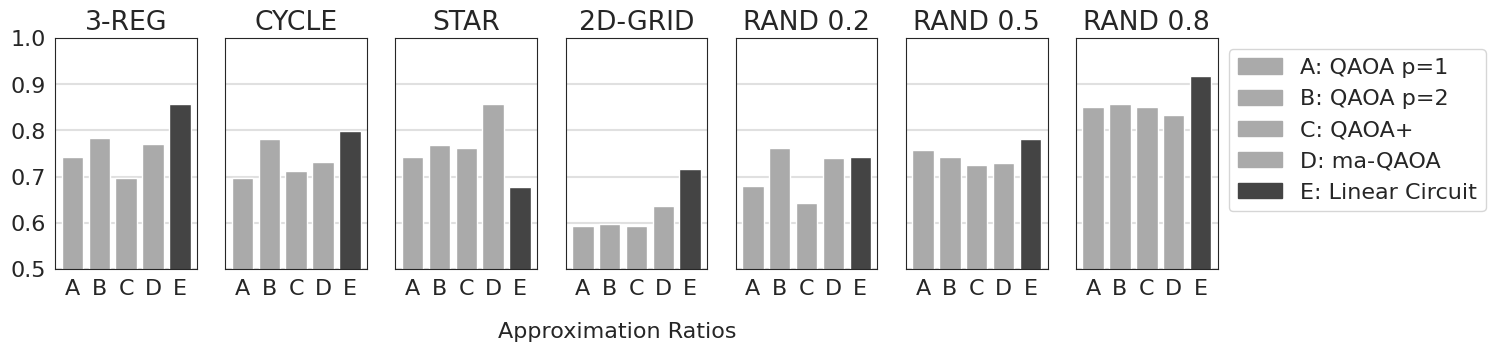}
    \caption{Approximation Ratio obtained by the tested quantum algorithms on Maximum Cut instances on graphs with 16 vertices.\label{fig:ar_maxcut}}
\end{figure*}

\figref{fig:ar_maxcut} represents the average approximation ratios obtained by executing the algorithms on the Maximum Cut problem across diverse graph topologies, as described above.
A notable observation is that the Linear circuit often achieves the best approximation ratio. However, there are exceptions, such as the Erdős–Rényi random graph with an edge probability of 0.2, where the Linear circuit performs slightly worse than QAOA with $p=2$. Moreover, on the star-topology graph, the approximation ratio obtained by the Linear circuit is significantly lower than that obtained by all other algorithms. We believe that further investigation into the cause of this behavior can be useful.

In contrast, when applying the same algorithms to Maximum Clique and Minimum Vertex Cover problems, we observe that the approximation ratios of the Linear circuit are lower than those achieved by the other algorithms. Among the algorithms tested, QAOA with $p=2$ consistently achieves the highest approximation ratio across all graph topologies, except for the Erdős–Rényi graphs with edge probabilities of 0.5 and 0.8, where QAOA with $p=1$ and ma-QAOA perform the best, respectively.
These findings support our observation that $R_{yz}$-connected ansatzes are suitable for problems where the cost of a solution remains the same if all bits are flipped, which include Maximum Cut, but not Maximum Clique and Minimum Vertex Cover problems.

\begin{figure}[ht]
    \centering
    \includegraphics[width=0.9\columnwidth]{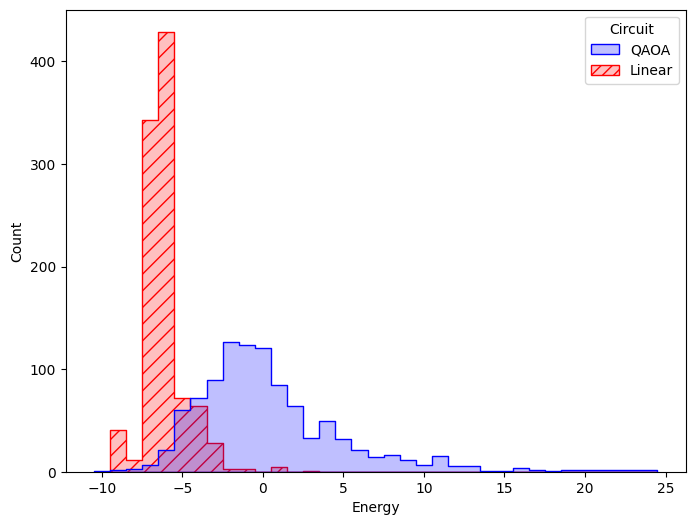}
    \caption{Comparison of solution distributions in circuits with optimal parameters solving the Maximum Cut problem on a $n=16$ vertices Erdős–Rényi graph with edge probability 0.8. Results are obtained using the Linear circuit and QAOA with $p=1$ with their parameter optimized, each executed with 1000 shots. The x-axis displays cost values of the QUBO formulation cost function, while the y-axis shows the frequency of each cost occurrence within the total shots. \label{fig:energy_distribution}}
\end{figure}

In order to better understand why the Linear ansatz achieves higher approximation ratios compared to other algorithms, we conduct an analysis of the final distribution of states obtained by the Linear circuit after its parameter optimization comparing it to QAOA with $p=1$ with 1000 shots, when applied to the Maximum Cut problem.
Specifically, we perform a single execution of both algorithms on the same problem instance and examine the distribution of solutions associated with the circuits with optimal parameters.
\figref{fig:energy_distribution} illustrates these distributions obtained on a Maximum Cut instance constructed from a $n=16$ vertices Erdős–Rényi graph with edge probability 0.5.
A notable observation is that the distribution of solutions produced by the Linear circuit is more concentrated on solutions with lower costs. This concentration is advantageous for achieving high approximation ratios according to our definition.
Conversely, the distribution of solutions generated by QAOA is more spread across the solution space, making this algorithm more suited for exploration purposes.

Overall, we can conclude that RLVQC was able to find a reasonable ansatz, consistently capable of finding higher approximation ratios than other quantum state-of-the-art algorithms.

\subsection{Implementability of $R_{yz}$-connected Ansatzes}
\label{subsec:implementability}
In this subsection we discuss the ease of implementability of the $R_{yz}$-connected ansatzes on specific real quantum devices.
In modern technologies, $R_z$ rotations can be implemented with negligible error and time for any rotation angle \cite{mckay_2017}, while implementing a $R_x$ rotation requires a specific application time.
Since any arbitrary rotation can be achieved by combining $R_z$ and $R_x(\frac{\pi}{2})$ rotations, superconducting computers are often calibrated to perform only $R_z$ and $R_x(\frac{\pi}{2})$ rotations, and not the full set of $R_x$ rotations.
Our $R_{yz}$-connected ansatzes are very good in such a framework, since we can observe that the $R_{yz}$ rotations can be decomposed into $R_z$ and $R_x(-\frac{\pi}{2})$ rotations, and $Cx$ gates.
In particular, the only single-qubit parametric gates are the $R_z$ rotations, which can be executed virtually without error.
Additionally, to implement the $R_x(-\frac{\pi}{2})$ rotations, we can substitute each of these gates with a $R_x(\frac{\pi}{2})$. $R_{yz}$-connected ansatzes maintain their performance after this substitution, allowing us to exploit the calibration of the hardware.

Moreover, the $R_{yz}$-connected family offers advantages when mapping the logic circuit to the quantum device topology. Specifically, we can select a particular element of the $R_{yz}$-connected ansatzes based on the connectivity of the available hardware. This approach reduces the number of SWAP gates required to connect qubits that are connected in the logical circuit but not in the hardware. This approach results in resource savings and reduces errors caused by an increased number of gates and circuit depth.

\section{Conclusions and Future Directions}
\label{sec:conclusions}

In this study, we introduced RLVQC, a Reinforcement Learning algorithm designed to learn how to construct quantum circuits to solve optimization problems through a quantum variational approach. 
Overall, RLVQC demonstrated its ability to build circuits able to generate solutions with approximation ratios comparable to those obtained by QAOA, and particularly high in the case of the Maximum Cut problem.

Furthermore, we analyzed the $R_{yz}$-connected ansatzes, a specific family of circuits discovered by the agent during its training on the Maximum Cut problem.
Our investigation revealed that the $R_{yz}$-connected ansatzes can achieve high approximation ratios, superior to those obtained with other state-of-the-art algorithms on Maximum Cut instances.

These results suggest that approaches employing Reinforcement Learning agents to construct variational circuits hold promise.
Given the flexibility of RL, there is potential for future research aimed at improving each of the components of RLVQC, including the representation of the environment's state, the agent's network architecture, the action set, and the reward function.
Each of these elements may be designed, for example, to take advantage of the peculiarities of a given problem, a specific hardware or even adapt to the requirements set by the task at hand.
More generally, RL looks promising for researchers aiming to address challenges characterized by large solution spaces within the quantum computing domain.

\section*{Acknowledgment}

We acknowledge the financial support from ICSC - ``National Research Centre in High Performance Computing, Big Data and Quantum Computing'', funded by European Union – NextGenerationEU.
We also acknowledge the support and computational resources provided by E4 Computer Engineering S.p.A.

\bibliographystyle{IEEEtran}
\bibliography{IEEEabrv,ms}

\end{document}